\documentclass[conference]{IEEEtran}
\makeatletter
\def\ps@headings{%
\def\@oddhead{\mbox{}\scriptsize\rightmark \hfil \thepage}%
\def\@evenhead{\scriptsize\thepage \hfil \leftmark\mbox{}}%
\def\@oddfoot{}%
\def\@evenfoot{}}
\makeatother
\usepackage[dvips]{graphicx}
\usepackage{graphics}
\usepackage{graphicx}
\usepackage{subfigure}
\usepackage{amssymb}
\usepackage{amsmath}
\usepackage{amsfonts}         
\usepackage{color}            
\usepackage{epic}             
\usepackage{eepic}            
\usepackage{rotating}         
\usepackage{type1cm}          
\usepackage{epsfig}
\usepackage{amsthm}
\usepackage{amsmath}
\usepackage{graphicx}
\usepackage{graphics}
\usepackage{epsfig}
\usepackage{latexsym}
\usepackage{amsfonts}
\usepackage{amssymb}
\usepackage{paralist}
\usepackage{xspace}
\usepackage{mathrsfs}
\usepackage{psfrag}
\usepackage{cite}
\usepackage{color}

\usepackage[ruled]{algorithm}
\usepackage[noend]{algorithmic}

\newtheorem{definition}{Definition}
\newtheorem{theorem}{Theorem}[section]

\def\auser{{\mathcal{U}}}
\def\aquery{{\mathcal{Q}}}
\def\asp{{\mathcal{P}}}

\def\homoEncrypt{{\mathbb E}}
\def\abeEncrypt{{\mathcal E}}

\begin{document}
\title{Search Me If You Can:\\Privacy-preserving Location Query Service}
\author{Xiang-Yang Li$^\mathcal{z}$ and Taeho Jung$^\mathcal{y}$\\
Department of Computer Science, Illinois Institute of Technology, Chicago, IL\\
$^\mathcal{z}$xli@cs.iit.edu, $^\mathcal{y}$tjung@hawk.iit.edu}

\maketitle
\footnotetext[1]{The research of Xiang-Yang Li is partially supported by NSF CNS-0832120, NSF CNS-1035894, NSF ECCS-1247944, National Natural Science Foundation of China under Grant No. 61170216, No. 61228202, China 973 Program under Grant No.2011CB302705.}

\begin{abstract}
Location-Based Service (LBS) becomes increasingly popular with the dramatic growth of smartphones and social network services (SNS), and its context-rich functionalities attract considerable users. Many LBS providers use users' location information to offer them convenience and useful functions. However, the LBS could greatly breach personal privacy because location itself contains much information. Hence, preserving location privacy while achieving utility from it is still an challenging question now. This paper tackles this non-trivial challenge by designing a suite of novel fine-grained Privacy-preserving Location Query Protocol (PLQP). Our protocol allows different levels of location query on encrypted location information for different users, and it is efficient enough to be applied in mobile platforms.
\end{abstract}

\IEEEpeerreviewmaketitle

\section{Introduction}
\label{sec:introduction}

Location Based Service (LBS) has become one of the most popular mobile applications due to the wide use of smartphones. The smartphones, equipped with GPS modules, have powerful computation ability to process holders' location information, and this brought the flood of LBS applications in the smartphone ecosystem. A good example is the smartphone camera: if one takes a photo with a smartphone camera, the location where the photo is taken is embedded in the picture automatically, which helps one's remembrance. Furthermore, the explosive growth of social network services (SNS) also assisted its growth by constructing connections between location information and social network. When a picture taken by a smartphone (location embedded) is uploaded to the Facebook album, the system automatically shows the location of the picture on the map, and this is shared with the owner's friends in the Facebook (unless the privacy setting specifies otherwise).

Many similar applications exploit both LBS and SNS. They offer several attractive functions, but location information contains much more information than barely the location itself, which could lead to unwanted information leakage. For example, when Alice and Bob both use check-in application in Facebook (which leaves a location record in one's webpage) in a nice restaurant, it is inferable that they are having a date and that they could be in a relationship. This inference might be an unintended information leakage from Alice's and Bob's perspective. Therefore, a privacy-preserving protocol is needed to prevent significant privacy breach resulted from the combination of LBS and SNS.

The simplest way, which most of applications adopted, is to exert group based access control on published locations: specify a group of user who can or cannot see them. Social photo sharing website Flickr only let users choose all users, neighbours, friends or family to allow the access to the locations, and SNS websites Facebook and Google+ additionally support custom groups to specify the accessible user groups. Mobile applications are much worse. Many mobile applications (e.g., Circle, Who's around and Foursquare) even do not offer group choices to the users, instead, they only ask users whether they want to disclose the location or not. Obviously, this is too simple to achieve what users need. First of all, from users' perspective, it is hard to explicitly determine a user group such that their locations are visible only to them. It is more natural to find a condition such that friends who satisfy it can or cannot see the location. Secondly, binary access control (can or cannot) is far beyond enough to properly configure the privacy setting. In the previous example of the two lovers Alice and Bob, Alice might want to share her date at the restaurant with her best friends and discloses the exact location to them. Besides, Alice might also want other friends to know that she is having a good time in downtown, but not detailed location. In this case, approximate settings between `can' and `cannot' are needed to fulfil her requirements.

As discussed above, existing privacy control settings in LBS are `coarse' in the sense that: 1) users can only explicitly specify a group of users who can or cannot access the location information; 2) access control policy supports binary choices only, which means users can only choose to enable or disable the information disclosure. The existing control strategies also suffer from privacy leakage in terms of the server storage. Even if one disables all of the location disclosure, his location is still open to the server, which in fact is users' top concern. Therefore, a fine-grained privacy control executable on encrypted location data is needed to further foster the LBS and its related business market.

\subsection{Contributions}
This paper proposes a fine-grained Privacy-preserving Location Query Protocol (PLQP) which enables queries to get location information (e.g., Searching a friend's approximate location, Finding nearest friends) without violating users location privacy. This is not a trivial job since simple anonymization makes it impossible to utilize them for queries. Also, if one directly applies queries or functions on the raw location information, privacy leakage is inevitable. Main contributions of our work are three-fold.
\begin{itemize}
\item \emph{Fine-Grained Access Control}:
Our protocol allows users to specify a condition instead of a group and exert access control over the users who satisfy this condition. This is more scalable since users can simply add a new condition for new privacy setting instead of hand-picking hundreds of users to form a new group. Also, this is more user-friendly because users themselves do not clearly know which of their friends should or should not access the information most of time.
\item \emph{Multi-leveled Access Control}:
The protocol also supports semi-functional encryption. That is, the protocol enables users to control to what extent (or level) others can learn his location. The lowest level corresponds to nothing, and the highest level corresponds to one's exact location. Levels between them correspond to indirect information about one's location.
\item \emph{Privacy-Preserving Protocol}:
In our protocol, every location information is encrypted and queries are processed upon ciphertexts. Therefore, a location publisher's friends learn nothing but the result of the location query, which is under the location publisher's control. In addition, since every location is encrypted, even the server who stores location information does not learn anything from the ciphertext.
\end{itemize}



\section{Related Work}
There are several works achieving privacy-preserving location query \cite{hashem2007safeguarding,bettini2005protecting,mokbel2006new,vuefficient}, which are based on $k$-anonymity model. The $k$-anonymity model \cite{sweeney2002k} has been widely used to protect data privacy. The basic idea is to remove some features such that each item is not distinguishable among other $k$ items. However, relevant techniques which achieve $k$-anonymity of data cannot be used in our case for the following four reasons: 1) Those techniques protect the privacy of the data stored in servers. In our PLQP, we do not store the data at all. 2) In LBS, location data is frequently updated, and this dynamic behaviour introduces huge overhead to keep the data $k$-anonymous. 3) As analyzed in Zang \textit{et al.} \cite{zang2011anonymization}, achieving $k$-anonymity in location dataset significantly violate the utility of it even for small $k$, so it is not suitable for our location query protocol. 4) $k$ is generally a system-wide parameter which determines the privacy level of all data in the system, but our goal is to leave the decision of privacy level to each user.

Kido \textit{et al.} \cite{kido2005protection} proposed a scheme which appends multiple false locations to a true one. The LBS responds to all the reports, and the client only collects the response corresponding to the true location. They examined this dummy-based technique and predicted how to make plausible dummy locations and how to reduce the extra communication cost. However, their technique protects the users' location privacy against LBS provider. We are also interested in a user's location privacy against other users.

In the mix zone model proposed by Beresford \textit{et al.} \cite{beresford2004mix}, users are assigned different pseudonyms every time he enters the mix zone, and users' paths are hidden by doing so. Several works \cite{hoh2008virtual,li2006swing,liutraffic} are based on this model, but they guarantee the privacy only when the user density is high and user behaviour pattern is unpredictable. Also, most of them require trusted servers.

There are also works related to CR (cloaking region) \cite{gedik2005location,kalnis2007preventing,wangl2p2,chow2006peer}. In these works, the LBS receives a cloaking region instead of actual users' locations. Gedit \textit{et al.} proposed spatial cloaking and temporal cloaking in \cite{gedik2005location}. Each query specifies a temporal interval, and queries within the same interval,  whose sources are in the vicinity of the first query's source, are merged to a single query. Otherwise, the query is rejected because it has no anonymity. Kalnis \textit{et al.} \cite{kalnis2007preventing} used the Hilbert space filling curve to map the two dimensional locations to one dimensional values, which are then indexed by a B+ tree. Then, they partition the one dimensional sorted list into groups of $n$ users, which is the CR of their scheme. Since this Hilbert Cloaking is not based on geometric space, it guarantees privacy for any location distribution. However, a certain range, where the user is located, is disclosed in CR-based approaches, and this is out of users' control. It is more desirable to allow users themselves to configure it.

\section{System Model and Problem Formulation}
\label{sec:model}

\subsection{System Model}
We denote every person engaged in the protocol as a user
$\auser_i$ (we do not differentiate smartphone users and PC
users), the user who publishes his location as a publisher
$\asp_i$ and the user who queries the location information of
other user as a querier $\aquery_i$.
Note that a user can be a querier and a publisher at the same time. When he queries on others, he acts as a querier and when he is queried, he acts as a publisher. That is, $\auser_i=\asp_i=\aquery_i$ for the same $i$.

Also, mobile applications or SNS applications which support LBS are denoted as service providers $\mathcal{SP}$. $\aquery$ and $\asp$ retrieves keys from $\mathcal{SP}$, which are used for access control. For simplicity, we consider only one $\mathcal{SP}$ here.

We assume an independent semi-honest model for users and service
providers. That is, they all behave independently and will try to
extract useful information from the ciphertexts, but they will follow
the protocol in general and will not collude with each other. We further assume that every user communicate
with each other via an anonymized network (e.g., Tor:
https://www.torproject.org) or other anonymized protocol (\cite{liu2011rumor}) such that the
privacy is not compromised by the underlying network protocol. We assume the origin of a packet is successfully hidden, which is out of this paper's scope (otherwise any attacker can achieve the location based on the origin of the packet).


\subsection{Location Assumption}

For simplicity, we assume the ground surface is a plane, and every
 user's location is mapped to an Euclidean space with integer
 coordinates (with meter as unit). That is, everyone's location can be
 expressed as a tuple of coordinates representing a point in a grid
 partition of the space.
This does not affect the generality since there exists a bijection
between spherical locations and Euclidean locations.
By approximating the coordinates in the Euclidean space   to the nearest
grid point,  we can show that it results in errors of the Euclidean
distance between two locations at most $\sqrt 2$ meters when the space
is partitioned using grid of side-length $1$ meter.

The Euclidean distance  between two users with locations
$\textbf{x}_1=(x_{11},x_{12},x_{13})$  and
$\textbf{x}_2=(x_{21},x_{22},x_{23})$ is
$dist(\auser_1,\auser_2)=|\textbf{x}_1-\textbf{x}_2|=\sqrt{\sum\limits_{i=1}^{3}{{(x_{1i}-x_{2i})}^2}}$.
Given a real location on the surface of the earth,
 we need to compute the surface distance, denoted as $SD(\auser_i,
 \auser_j)$, between these two points.
By assuming that the earth is a sphere with radius $R$ meters,
 it is easy to show that
 $SD(\auser_i, \auser_j)= 2 \arcsin(\frac{dist(\auser_i, \auser_j)}{2R})
 \cdot R $.
Then the surface distance can be quickly computed from the Euclidean
distance.
To check if the surface distance satisfies certain conditions, we can
convert it to check if the Euclidean distance satisfying corresponding
conditions.
For example, $dist(\auser_1,\auser_2) \le D$ is equivalent as
 $SD(\auser_i, \auser_j) \le 2R \arcsin(D /2R)$.
For simplicity and convenience of presentation, in this paper, we will
 focus on the Euclidean distance instead of the surface distance.
Notice that although
 we consider only Euclidean space here, our protocol works for
 any system where distance is a polynomial of location points
 \textbf{x}'s, where \textbf{x} is a vector.

\subsection{Problem Statement}
Each user $\auser_i$ has his location information
$\textbf{x}_i=(x_{i1},x_{i2},x_{i3})$ which determines his current
location.
He also has an attribute set $S_i$ which determines his identity
(e.g., University:I.I.T, Degree:Ph.D, Major:Computer Science). Then, a
querier $\aquery_i$ uses his current location information and
attribute set to execute a query (function) $f$ on a publisher
$\asp_j$'s location information $\textbf{x}_j$. According to
$\aquery_i$'s location information $\textbf{x}_i$ and his attribute
set $S_i$, he obtains the corresponding query result
$f(\textbf{x}_i,S_i,\textbf{x}_j)$. Note that different $\textbf{x}_i$
and $S_i$ leads to different level of query result. During the whole
protocol, $\aquery_i$ or $\asp_j$ cannot learn any useful extra
information about each other's location information.

In this paper, we propose novel protocols such that the location
publisher exerts a fine-grained access control on who can access what location
information.
For example, a publisher could specify the following access control
policies:
(1) a user can know which city I am in if s/he is in my friend list;
or (2) a user can check whether the distance between him and me is
less than 100 meters if s/he is my classmate;
 or (3) a user can compute the exact distance between us if we both
 went to the same university.
We generally assume that a user $\auser_i$ has a set of attributes
$A_i$, and that an access control policy of the publisher is specified
by a boolean
function (specified as an access tree $T$) on all possible attributes
of users.

According to the location information disclosed to the querier,
 we define four different levels of queries.
\begin{definition}
\label{def:level-1}
\emph{Level 1 Query}: When the query ends, $\aquery$ learns whether
$dist(\aquery,\asp)\le \tau$ or not if the attributes of
the querier satisfy a certain condition specified by the publisher,
where $\tau$ is a threshold  value determined by $\asp$.
The querier knows nothing else about the location of the publisher.
\end{definition}

\begin{definition}
\label{def:level-2}
\emph{Level 2 Query}: When the query ends, $\aquery$ learns whether
$dist(\aquery,\asp)\le \tau$ when the attributes of the querier
satisfy a certain condition specified by the publisher,
 where $\tau$ is a threshold value determined by $\aquery$.
The querier knows nothing else about the location of the publisher.
\end{definition}

\begin{definition}
\label{def:level-3}
\emph{Level 3 Query}: When the query ends, $\aquery$ learns the
$dist(\aquery,\asp)$ if the attributes of the querier
satisfy a certain condition specified by the publisher.
The querier knows nothing else about the location of the publisher.
\end{definition}

\begin{definition}
\label{def:level-4}
\emph{Level 4 Query}: When the query ends, $\aquery$ learns the 
function $F(\textbf{x})$ of the location $\textbf{x}$ of $\asp$ if the
attributes of the querier
satisfy a certain condition  specified by the publisher. Here function
$F$ is defined by the publisher.
The querier knows nothing else about the location of the publisher.
\end{definition}

It is easy to show that the level $i$ query provides better privacy
protection than level $i+1$ query, for $i=1,2$.
Level $4$ query provides most information in general.
In level 4 query, the function $F$ could be used by the publisher to
exert fine-grained access control on his location information.
For example $F(\textbf{x})$ could return the city of the location, 
the zip-code of the location or the exact location information.


\section{Background}
\label{sec:background}

In our Privacy-preserving Location Query Protocol (PLQP), various cryptographic concepts are used. We introduce each of them in this section.

\subsection{Attribute-Based Encryption (ABE)}
As Jung \textit{et al.} discussed in detail in their work
\cite{jung2013cloud}, in the Attribute-Based Encryption (ABE)
\cite{abe}, the identity of a person is viewed as a set of attributes. This enables the encrypter to specify a boolean function to do access control. There are two types of ABE system: Goyal \textit{et al.}'s Key-Policy Attribute-Based Encryption \cite{kp-abe} and Bethencourt \textit{et al.}'s Ciphertext-Policy Attribute-Based Encryption \cite{cp-abe}. The KP-ABE specifies the encryption policy in the decryption key, and the CP-ABE specifies the policy in the ciphertext. Due to many reasons discussed in \cite{jung2013cloud}, we will employ CP-ABE as a component of access control.

\subsubsection{Access Tree $T$}
In most of previous ABE works (e.g.,\cite{kp-abe}\cite{cp-abe}\cite{hasbe}), encryption policy is described with an access tree. Each non-leaf node of the tree is a threshold gate by a threshold value $\theta$, and each leaf node $x$ is described by an attribute. A leaf node is satisfied if a key contains the corresponding attribute, and a non-leaf threshold gate is satisfied if at least $\theta$ children are satisfied.

Note that this threshold-gate based access tree is able to express arbitrary condition, which makes the privacy control in our protocol flexible and scalable.

\subsubsection{Definition}
With the access tree defined as above, the CP-ABE scheme is defined as follows:
\vspace{8pt}

\textsf{Setup $\rightarrow \mathbf{PK}, \mathbf{MK}$}.   \hspace{4pt}The setup algorithm takes nothing as input other than the implicit security parameter. It outputs the public parameter $\mathbf{PK}$ and a master key $\mathbf{MK}$. The master key belongs to the key issuer and is kept secret.
\vspace{8pt}

\textsf{Encrypt($\mathbf{PK}$, $M$, $T$) $\rightarrow \abeEncrypt_{T}(M)$}.   \hspace{4pt}The encryption algorithm takes as input the public key $\mathbf{PK}$, a message $M$, and an access tree $T$. It will encrypt the message $M$ and returns a ciphertext $\mathbf{CT}$ such that only a user with key satisfying the access tree $T$ can decrypt it.
\vspace{8pt}

\textsf{KeyGenerate($\mathbf{PK}$, $\mathbf{MK}$, $S$) $\rightarrow \mathbf{SK}$}.   \hspace{4pt}The Key Generation algorithm takes as input the public key $\mathbf{PK}$, the master key $\mathbf{MK}$ and a set of attributes $S$. It outputs a private key $\mathbf{SK}$ which contains the attributes in $S$.
\vspace{8pt}

\textsf{Decrypt($\mathbf{PK}$, $\mathbf{SK}$, $\abeEncrypt_{T}(M)$) $\rightarrow M$}.   \hspace{4pt}The decryption algorithm takes as input the public parameter $\mathbf{PK}$, a private key $\mathbf{SK}$ whose attribute set is $S$, and a ciphertext $\mathbf{CT}$ which contains an access tree $T$. It outputs the original message $M$ if and only if the set $S$ satisfies the access tree $T$.

We direct the readers to \cite{cp-abe} for detailed construction.

\subsection{Homomorphic Encryption (HE)}
Homomorphic Encryption (HE) allows direct addition and multiplication on ciphertexts while preserving decryptability. That is, following equations are satisfied (Note that this is only an example of a HE, and detailed operations vary for different HE system). 
\begin{displaymath}
\text{Enc}(m_1)\cdot \text{Enc}(m_2)=\text{Enc}(m_1+m_2)
\end{displaymath}
\begin{displaymath}
\text{Enc}(m_1)^{\text{Enc}(m_2)}=\text{Enc}(m_1\cdot m_2)
\end{displaymath}
where $\text{Enc}(m)$ stands for the ciphertext of $m$. 

In general, there are two types of HE: Partially Homomorphic Encryption (PHE) and Fully Homomorphic Encryption (FHE). PHE supports constant number of additions and multiplications, and FHE supports unlimited additions and multiplications but it is much less efficient than PHE. As discussed by Lauter \textit{et al.} in \cite{lauter2011can}, the decryption time of FHE system is too high to be used in a real application, and in most of cases one only needs a few number of multiplications or additions. Therefore, Pallier's system, which is much simpler and thus efficient, is our choice: it involves only one multiplication for each homomorphic addition and one exponentiation for each homomorphic multiplication.

\subsubsection{Definition of Paillier's Cryptosystem}\label{section:paillier}
Paillier's cryptosystem is composed of three algorithms -- \textsf{KeyGenerate}, \textsf{Encrypt} and \textsf{Decrypt}.
\vspace{8pt}

\textsf{KeyGenerate} $\rightarrow EK, DK$.  An entity randomly chooses two large prime numbers $p$ and $q$ of same bit length. He then computes $n=pq$ and $\lambda=(p-1)(q-1)$. Next, he sets $g=(n+1)$ and $\mu=(\lambda \text{ mod }n^2)^{-1}$ mod $n$. Then, the encryption key is $EK=(n,g)$ and the decryption key is $DK=(\lambda, \mu)$.
\vspace{8pt}

\textsf{Encrypt}$(EK, m)$ $\rightarrow \homoEncrypt(m,r)$.    The encrypter selects a random integer $r\in\mathbb{Z}_n$ and computes the ciphertext
\begin{displaymath}
\homoEncrypt(m,r)=g^m\cdot r^n \text{ mod } n^2
\end{displaymath}
and publishes it.
\vspace{8pt}

\textsf{Decrypt}$(\homoEncrypt(m,r),DK)$ $\rightarrow m$.   The holder of $DK=(\lambda, \mu)$ can decrypt the ciphertext $\homoEncrypt(m,r)$. He computes the following to recover the message:
\begin{displaymath}
m=L((\homoEncrypt(m,r))^{\lambda} \text{ mod }n^2)\cdot\mu \text{ mod }n
\end{displaymath}
where $L(a)=(a-1)/n$ mod $n$.

The Paillier's cryptosystem satisfies the following homomorphic properties:
\begin{displaymath}
\homoEncrypt(m_1,r_1)\cdot \homoEncrypt(m_2,r_2)=\homoEncrypt(m_1+m_2,r_1r_2) \text{ mod }n^2
\end{displaymath}
\begin{displaymath}
\homoEncrypt(m_1,r_1)^{m_2} = \homoEncrypt(m_1\cdot m_2,r_1^{m_2})\text{ mod }n^2
\end{displaymath}

Note that $DK$ can decrypt only the ciphertexts encrypted with $EK$ which pairs with it. Also, the random number $r$ in a ciphertext $\homoEncrypt(m,r)$ does not contribute to decryption or other homomorphic operation. It only prevents the dictionary attack by randomizing the ciphertext. For sake of simplicity, we use $\homoEncrypt(m)$ instead of $\homoEncrypt(m,r)$ in the remaining paper.

\subsection{Functional Encryption (FE)}
Functional Encryption (FE) is a new encryption scheme recently proposed after the Attribute-Based Encryption (ABE). To the best of our knowledge, the concept is first proposed by Boneh \textit{et al.} in \cite{boneh2008impossibility}. In the open direction of their work, they proposed the terminology `Functional Encryption' and its general concept, and later in 2011, Boneh \textit{et al.} formally defined it and discussed its challenge \cite{boneh2011functional}. According to their study, the FE is defined as follows: FE is an encryption scheme such that a key holder can learn a specific function of the data based on the ciphertext, but nothing else about the data. This is totally different from the traditional encryption scheme in terms of the differentiated decryption. In traditional encryption schemes (e.g., PKI, ABE), decryption result of a ciphertext for every authorized users is same: the plaintext. In FE, encrypter can specify a function for each key such that each decryption result is the corresponding function of the plaintext.

There are a few recent works related to FE (\cite{lewko2010fully,sahai2010worry}). However, they mainly focus on hiding encryption policy from ordinary users. To the best of our knowledge, there is no formal construction of FE which satisfies the definition of FE \cite{boneh2011functional}.

\section{Preliminary Design}
\label{sec:prelim-design}

In our PLQP, we require that a publisher could specify several access
control structures for all potential location queriers.
Different access trees will allow access to different
level of knowledge about the location information, which
 is achieved by using FE in our protocol.
However, strictly speaking, the encryption in our protocol is not a
 formal FE because we
 only support a constant number of functions of the data, so we refer
 to it as semi-functional encryption.
To allow a set of possible queries by all users, we first present
  distance computation and comparison algorithms which will be used
 to provide four levels of functions over location data in our
 semi-functional PLQP.

\subsubsection{Privacy Preserving Distance
  Computation}\label{section:computation}

Let $\textbf{x}=(x_1,x_2,x_3)$ and $\textbf{y}=(y_1,y_2,y_3)$ be a
publisher $\asp$'s and a querier $\aquery$'s 3-dimensional
location respectively. We use Algorithm \ref{alg:distance-computing} to let $\aquery$ securely compute $dist(\asp,\aquery)$ without knowing
$\asp$'s coordinates or disclosing his own one.

\begin{algorithm}[hptb]
\caption{Privacy Preserving Distance Computation}
\label{alg:distance-computing}
\begin{algorithmic}[1]

\STATE $\aquery$ generates a pair of encryption and decryption
keys of Paillier's cryptosystem: $EK=(n,g)$, $DK=(\lambda,\mu)$.
 We assume $n$ is of 1024-bit length.
$\homoEncrypt_{\aquery}$ denotes the encryption done by $\aquery$ using his
 encryption keys.

\STATE $\aquery$ generates the following ciphertexts and sends
them to $\asp$ at $\textbf{x}$.
\[{\small \homoEncrypt_{\aquery}(1),\homoEncrypt_{\aquery}(\sum_{i=1}^{3}y_i^2),\{\homoEncrypt_{\aquery}(y_i)
\mid  i=1,2,3\},}
\]

\STATE $\asp$, after receiving the ciphertexts, executes the following
homomorphic operations:
\begin{eqnarray*}
\begin{cases}
&\{\homoEncrypt_{\aquery}(y_i)^{-2x_i}\}=\{\homoEncrypt_{\aquery}(-2x_iy_i)\}, \mbox{ for } i=1,2,3 \\
&\homoEncrypt_{\aquery}(1)^{\sum_{i=1}^3 x_i^2}= \homoEncrypt_{\aquery}(\sum_{i=1}^{3}x_i^2)
\end{cases}
\end{eqnarray*}

\STATE $\asp$ computes and sends the following  to the querier $\aquery$:
\[{\small \begin{split}
&\homoEncrypt_{\aquery}(\sum_{i=1}^{3}x_i^2)\cdot
  \homoEncrypt_{\aquery}(\sum_{i=1}^{3}y_i^2)\cdot
  \prod_{i=1}^{3}(\homoEncrypt_{\aquery}(-2x_iy_i))\\
&=\homoEncrypt_{\aquery}(\sum_{i=1}^{3}(x_i-y_i)^2)
=\homoEncrypt_{\aquery}(|\textbf{x}-\textbf{y}|^2)
\end{split}}\]

\STATE $\aquery$ uses the private key $DK$ to decrypt the
$\homoEncrypt_{\aquery}(|\textbf{x}-\textbf{y}|^2)$ to get the distance.
\end{algorithmic}
\end{algorithm}

Note that the location \textbf{y} is kept secret to $\asp$ during the
whole protocol, since he does not know the private key; on the other
hand, the location \textbf{x} is also kept secret since $\aquery$ only
achieves $\homoEncrypt(|\textbf{x}-\textbf{y}|^2)$. However, the location
\textbf{x} is inferred if $\aquery$ runs the same protocol at
different places for four times in Euclidean space (three times in
Euclidean plane). This will be discussed in detail in Theorem
\ref{theorem:3}.

\subsubsection{Privacy Preserving Distance Comparison}\label{section:compare}

Let $\textbf{x}=(x_1,x_2,x_3)$ and $\textbf{y}=(y_1,y_2,y_3)$ be
publisher $\asp$'s and querier $\aquery$'s 3-dimensional location
respectively. We use Algorithm \ref{alg:secure-comparison} to let $\aquery$ learn
whether $dist(\asp,\aquery)$ is less than, equal
to or greater than a threshold value $\tau$, which is determined by
the publisher $\asp$.

\begin{algorithm}[hptb]
\caption{Privacy Preserving Distance Comparison}
\label{alg:secure-comparison}
\begin{algorithmic}[1]
\STATE
 $\aquery$ generates encryption and decryption key pair of Paillier's
cryptosystem: $EK=(n,g),DK=(\lambda,\mu)$.

\STATE $\aquery$ generates the following ciphertexts and sends them to the user $\asp$ with location
\textbf{x}.
\[ \small{
\homoEncrypt_{\aquery}(1),\homoEncrypt_{\aquery}(\sum_{i=1}^3 y_i^2),\{\homoEncrypt_{\aquery}(-2y_i)  \mid i=1,2,3 \}}
\]

\STATE $\asp$, after receiving the ciphertexts, randomly picks two integers $\delta\in\mathbb{Z}_{2^{972}},\delta'\in\mathbb{Z}_{2^{1022}}$ and executes the following homomorphic operations:
{\small \begin{eqnarray*}
\begin{cases}
& \{\homoEncrypt_{\aquery}(-2y_i)^{\delta x_i}=\homoEncrypt_{\aquery}(-2\delta x_iy_i)\mid i=1,2,3 \} \\
& \homoEncrypt_{\aquery}(\sum_{i=1}^{3}y_i^2)^{\delta}=\homoEncrypt_{\aquery}(\delta(y_1^2+y_2^2+y_3^2))\\
& \homoEncrypt_{\aquery}(1)^{\delta\sum_{i=1}^3x_i^2}=\homoEncrypt_{\aquery}(\delta \sum_{i=1}^3x_i^2))\\
&\homoEncrypt_{\aquery}(1)^{\delta'}=\homoEncrypt_{\aquery}(\delta') \\
&\homoEncrypt_{\aquery}(\delta \sum_{i=1}^3 x_i^2)\cdot \homoEncrypt_{\aquery}(\delta')
 =\homoEncrypt_{\aquery}(\delta\sum_{i=1}^3 x_i^2+\delta')
\end{cases}
\end{eqnarray*}
}

\STATE $\asp$ computes the followings and sends them back to the other user at \textbf{y}.
{\small
\[ \begin{split}
&\homoEncrypt_{\aquery}(\delta \sum_{i=1}^3 x_i^2+\delta')\cdot \homoEncrypt_{\aquery}(\delta \sum_{i=1}^3 y_i^2)
\prod_{i=1}^{3}(\homoEncrypt_{\aquery}(-2\delta x_iy_i))\\
&=\homoEncrypt_{\aquery}((\delta\sum_{i=1}^{3}(x_i-y_i)^2)+\delta')=\homoEncrypt_{\aquery}(\delta|\textbf{x}-\textbf{y}|^2+\delta') \\
& \homoEncrypt_{\aquery}(1)^{\delta\tau^2+\delta'}=\homoEncrypt_{\aquery}(\delta\tau^2+\delta')
\end{split} \]
}

\STATE $\aquery$ uses the private key $DK(\lambda,\mu)$ to decrypt the ciphertexts and gets  $\delta|\textbf{x}-\textbf{y}|^2+\delta'$ and $\delta\tau^2+\delta'$. If, without modular operations, both of them are less than the modulo $n$, we have:
\begin{displaymath}
\delta|\textbf{x}-\textbf{y}|^2+\delta'<\delta\tau^2+\delta' \ \ 
\Leftrightarrow \ \  |\textbf{x}-\textbf{y}|<\tau
\end{displaymath}

\end{algorithmic}
\end{algorithm}

The reason $\delta$ and $\delta'$ are chosen from $\mathbb{Z}_{2^{972}}$ and $\mathbb{Z}_{2^{1022}}$ is because otherwise the comparison is not correct due to the modular operations. This will be further discussed in Section \ref{section:deltas}.

On the other hand, if $\aquery$ wants to determine the threshold value $\tau$, he can send another ciphertext $\homoEncrypt(\tau^2)$ at the Step 2. Then, $\asp$ computes $\homoEncrypt(\tau^2)^\delta\cdot \homoEncrypt(1)^{\delta'}=\homoEncrypt(\delta\tau^2+\delta')$ at the Step 4 and proceeds same as Algorithm \ref{alg:secure-comparison}.

\section{Privacy Preserving Location Services}
\label{section:construction}

In this section, we propose the construction of Privacy-preserving Location Query Protocol (PLQP). First of all, we define a group for CP-ABE.

Let $\mathbb{G}_0$ be a multiplicative cyclic group of prime order $m$ and $g$ be its generator. The bilinear map $e$ used in CP-ABE is defined as follows:  $e:\mathbb{G}_0\times\mathbb{G}_0\rightarrow\mathbb{G}_T$, where $\mathbb{G}_T$ is the codomain of the map $e$. The bilinear map $e$ has the following properties:
\begin{compactenum}
\item \textbf{Bilinearity}:
for all $u,v\in \mathbb{G}_0$ and $a,b\in \mathbb{Z}_q$, $e(u^a,v^b)=e(u,v)^{ab}$
\item \textbf{Symmetry}:
for all $u,v\in \mathbb{G}_0$, $e(u,v)=e(v,u)$
\item
\textbf{Non-degeneracy}:
$e(g,g)\neq 1$
\end{compactenum}

\begin{definition}
The Decisional Diffie-Hellman (DDH) problem in an integer group with generator $g$ is defined as follows: on input $g, g^a, g^b, g^c=g^{ab}\in\mathbb{Z}$, where $a,b,c\in \mathbb{Z}$, decide whether $c=ab$ or $c$ is a random element.
\end{definition}

\begin{definition}
The Decisional Bilinear Diffie-Hellman (DBDH) problem in group $\mathbb{G}_0$ of prime order $p$ with generator $g$ is defined as follows: on input $g, g^a, g^b, g^c\in\mathbb{G}_0$ and $e(g,g)^{z}=e(g,g)^{abc}\in\mathbb{G}_T$, where $a,b,c\in \mathbb{Z}_q$, decide whether $z=abc$ or $z$ is a random element.
\end{definition}

The security of our construction relies on the assumption that no probabilistic polynomial-time algorithms can solve the DDH problem or DBDH problem with non-negligible advantage. This is a widely made assumption in various cryptographic works (\cite{jung2013cloud,boneh2001identity,sahai2010worry}), which is reasonable since discrete logarithm problems in large number fields are widely considered to be intractable (\cite{jung2013cloud,jung2013data,zhang2013verifiable}).

The reason we introduce CP-ABE here is to exert fine-grained access control over the location queries. Even if one's location satisfies a certain condition, one cannot gain any information from the query if his identity attributes do not satisfy the pre-defined encryption policy. 

\subsection{Initialize}
The service provider $\mathcal{SP}$ initializes the system by following the instructions:

\begin{algorithm}[ht]
\caption{Initialize}
\begin{algorithmic}[1]
\STATE
 Executes \textsf{Setup} (CP-ABE) to generate public and master key pairs:
\begin{displaymath}
\begin{cases}
\mathbf{PK}=\langle \mathbb{G}_0,g,h=g^\beta,f=g^{1/\beta},e(g,g)^\alpha\rangle \\
\mathbf{MK}\langle \beta,g^\alpha\rangle
\end{cases}
\end{displaymath}

\STATE
 Executes \textsf{KeyGenerate} (CP-ABE) for all users within the system to issue them private keys corresponding to their attributes.
\begin{displaymath}
\mathbf{SK}=\langle D=g^{(\alpha+r)/\beta},\forall j\in S:~D_j=g^r H(j)^{r_j},D'_j=g^{r_j}\rangle
\end{displaymath}
\end{algorithmic}
\end{algorithm}

Here we assume secure channels exist between users and service providers $\mathcal{SP}$ such that
private keys are securely delivered to each user.

\subsection{Protocol for Level 4 Query}
After the query ends, $\aquery_j$ learns $\asp_i$'s exact location $\textbf{x}_i$.

\begin{algorithm}[ht]
\caption{Level 4 Query Protocol}
\begin{algorithmic}[1]
\STATE
 A publisher $\asp_i$ creates an access tree $T_{i4}$ which specifies the access authority for the level 4 query.

\STATE
 When a querier $\aquery_j$ sends a level 4 query to $\asp_i$, $\asp_i$ encrypts his location using the CP-ABE algorithm \textsf{Encrypt}:
\begin{displaymath}
\abeEncrypt_{T_{i4}}(x_{i,1}),\abeEncrypt_{T_{i4}}(x_{i,2}),\abeEncrypt_{T_{i4}}(x_{i,3})
\end{displaymath}

\STATE
 These are sent to $\aquery_j$, and $\aquery_j$ decrypts it with his private key $\mathbf{SK}$ if it satisfies the access tree $T_{i4}$, and achieves $\asp_i$'s locoation.
\end{algorithmic}\end{algorithm}

\subsection{Protocol for Level 3 Query}
After the query ends, $\aquery_j$ learns the $dist(\aquery_j,\asp_i)$.

\begin{algorithm}[ht]
\caption{Level 3 Query Protocol}
\begin{algorithmic}[1]
\STATE A publisher $\asp_i$ creates an access tree $T_{i3}$ which specifies the access authority for the level 3 query.

\STATE When a querier $\aquery_j$ wants to send a level 3 query to $\asp_i$, he initiates the Secure Distance Computation protocol (Section \ref{section:computation}) by generating encryption and decryption Paillier key pair $EK_j=(n_j,g_j),DK_j=(\lambda_j,\mu_j)$.

\STATE Then, he calculates the following ciphertexts and sends to $\asp_i$:
\begin{displaymath}
\homoEncrypt(1),\homoEncrypt(x_{j1}^2+x_{j2}^2+x_{j3}^2),\{\homoEncrypt(-2x_{ji})\}_{i=1,2,3}
\end{displaymath}

\STATE
 $\asp_i$, after receiving them, calculates the ciphertext below:
\begin{displaymath}
\homoEncrypt(|\textbf{x}_i-\textbf{x}_j|^2)
\end{displaymath}

\STATE
 The ciphertext above is encrypted again with the access tree $T_{i3}$ using the CP-ABE algorithm \textsf{Encrypt},  which we refer to doubly nested ciphertexts:
\begin{displaymath}
\abeEncrypt_{T_{i3}}(\homoEncrypt(|\textbf{x}_1-\textbf{x}_2|^2)
\end{displaymath}

 \STATE
 The doubly nested ciphertext is sent back to $\aquery_j$, and if $\aquery_j$'s private key $\mathbf{SK}$ satisfies the access tree $T_{i3}$, he can decrypt it and use his Paillier key pair to decrypt the ciphertext again to achieve $|\textbf{x}_1-\textbf{x}_2|^2$. Then, he obtains the $dist(\aquery_j,\asp_i)$.
\end{algorithmic}\end{algorithm}

\begin{theorem}
\label{theorem:3}
If $\aquery$ executes the level 3 query for more than three times at different places in Euclidean space, level 3 query is equivalent to level 4 query.
\end{theorem}
\begin{proof}
This is also mentioned in the Section \ref{section:compare}. If $\aquery$ executes the level 3 query for four times at different locations, he achieves 4 distances: $\{|\textbf{x}_{i}-\textbf{y}|\}_{i=1,2,3,4}$, where $\textbf{x}_i$'s are $\aquery$'s 4 different locations and $\textbf{y}$ is $\asp$'s location. These are essentially four equations with three variables $y_1,y_2$ and $y_3$:
\begin{displaymath}
(x_{i1}-y_1)^2+(x_{i2}-y_2)^2+(x_{i3}-y_3)^2=|\textbf{x}_i-\textbf{y}|^2~~(i=1,2,3,4)
\end{displaymath}
which can be solved. Therefore, $\asp$'s location \textbf{y} can be computed in this case.
\end{proof}

Similarly, it can be proved that if $\aquery$ executes the level 3 query for more than two times at different places in Euclidean plane, level 3 query is equivalent to level 4 query.

\subsection{Protocol for Level 2 Query}

After the query ends, $\aquery_j$ learns whether $dist(\aquery_j,\asp)$ is less than, equal to or greather than $\tau$, where $\tau$ is a threshold value determined by $\aquery_j$.

\begin{algorithm}[ht]
\caption{Level 2 Query Protocol}
\begin{algorithmic}[1]
\STATE
 A publisher $\asp_i$ creates an access tree $T_{i2}$ which specifies the access authority for the level 2 query.
\STATE When a querier $\aquery_j$ wants to send a level 2 query to $\asp_i$, he initiates the Secure Distance Comparison protocol (Section \ref{section:compare}) by picking two large prime numbers $p_j,q_j$ of the same length, then $n_j=p_jq_j$, $g_j=n_j+1$, $\lambda_j=(p_j-1)(q_j-1)$ and $\mu_j=\lambda_j^{-1}$ mod $n_j$, which form Paillier key pair $EK_j=(n_j,g_j),DK_j=(\lambda_j,\mu_j)$ (The subscriptions indicate that these keys are used by $\aquery_j$).

\STATE
 Then, he calculates the following ciphertexts and sends them to $\asp_i$:
\begin{displaymath}
\homoEncrypt(1),\homoEncrypt(x_{j1}^2+x_{j2}^2+x_{j3}^2),\{\homoEncrypt(-2x_{ji})\}_{i=1,2,3}, \homoEncrypt(\tau^2)
\end{displaymath}

\STATE
 $\asp_i$, after receiving them, picks two random integers $\delta\in\mathbb{Z}_{2^{970}},\delta'\in\mathbb{Z}_{2^{1022}}$ and calculates the ciphertexts:
{\small \[
\begin{cases}\begin{split}
&\homoEncrypt(\sum_{k=1}^3x_{j,k}^2)^\delta\cdot \homoEncrypt(1)^{\delta \sum_{k=1}^3x_{i,k}^2+\delta'} \cdot\prod_{k=1}^{3}{\homoEncrypt(-2x_{jk})^{\delta x_{ik}}}\\
& \homoEncrypt(\tau^2)^{\delta}\cdot \homoEncrypt(1)^{\delta'}=\homoEncrypt(\delta\tau^2+\delta')
\end{split}\end{cases}
\]}
where $\textbf{x}_i$ and $\textbf{x}_j$ refer to $\asp_i$'s and $\aquery_j$'s locations respectively.

\STATE
 These ciphertexts are encrypted again with the access tree $T_{i1}$ using the CP-ABE algorithm \textsf{Encrypt}:
\begin{displaymath}
\abeEncrypt_{T_{i1}}(\homoEncrypt(\delta|\textbf{x}_i-\textbf{x}_j|^2+\delta')), \abeEncrypt_{T_{i1}}(\homoEncrypt(\delta\tau^2+\delta'))
\end{displaymath}

\STATE The doubly nested ciphertexts are sent back to $\aquery_j$, and if $\aquery_j$'s private key $\mathbf{SK}$ satisfies the access tree $T_{i1}$, he can decrypt them and uses his Paillier key pair to decrypt the ciphertext again to achieve $\delta|\textbf{x}_i-\textbf{x}_j|^2+\delta'$ and $\delta\tau^2+\delta'$. Then he is able to compare two values to learn whether $dist(\asp_i,\aquery_i)$ is less than, equal to or greater than $\tau$.
\end{algorithmic}\end{algorithm}

\begin{theorem}\label{theorem:2}
Suppose $D$ is the greatest possible distance in the location space, if $\aquery$ executes the level 2 query for $\Theta(\log D)$ times, level 2 query is equivalent to level 3 query.
\end{theorem}
\begin{proof}
Since $\aquery$ can control the threshold value $\tau$, he can first execute a level 2 query with $\tau=D$. Then, he uses binary search to execute level 2 queries with different $\tau$'s until he finds the $\tau$ such that $\tau=|\textbf{x}-\textbf{y}|$, where $\textbf{x}$ and $\textbf{y}$ are $\aquery$'s and $\asp$'s locations respectively. Then, he finds the distance.
\end{proof}

\subsection{Protocol for Level 1 Query}
After the query ends, $\aquery_j$ learns whether $dist(\aquery_j,\asp)$ is less than, equal to or greather than $\tau$, where $\tau$ is a threshold value determined by $\asp_i$.

\begin{algorithm}[ht]
\caption{Level 1 Query Protocol}
\begin{algorithmic}[1]
\STATE A publisher $\asp_i$ creates an access tree $T_{i1}$ which specifies the access authority for the level 1 query.

\STATE When a querier $\aquery_j$ wants to send a level 1 query to $\asp_i$, he initiates the Secure Distance Comparison protocol (Section \ref{section:compare}) by picking two large prime numbers $p_j,q_j$ of the same length, then $n_j=p_jq_j$, $g_j=n_j+1$, $\lambda_j=(p_j-1)(q_j-1)$ and $\mu_j=\lambda_j^{-1}$ mod $n_j$, which form Paillier key pair $EK_j=(n_j,g_j),DK_j=(\lambda_j,\mu_j)$ (The subscriptions indicate that these keys are used by $\aquery_j$).

\STATE Then, he calculates the following ciphertexts and sends them to $\asp_i$:
\begin{displaymath}
\homoEncrypt(1),\homoEncrypt(x_{j1}^2+x_{j2}^2+x_{j3}^2),\{\homoEncrypt(-2x_{ji})\}_{i=1,2,3}
\end{displaymath}

\STATE $\asp_i$, after receiving them, picks two random integers $\delta\in\mathbb{Z}_{2^{970}},\delta'\in\mathbb{Z}_{2^{1022}}$ and calculates the ciphertexts:
\begin{displaymath}
\homoEncrypt(\delta|\textbf{x}_i-\textbf{x}_j|^2+\delta'), \homoEncrypt(\delta\tau^2+\delta')
\end{displaymath}
where $\textbf{x}_i$ and $\textbf{x}_j$ refer to $\asp_i$'s and $\textbf{x}_j$'s locations respectively.

\STATE These ciphertexts are encrypted again with the access tree $T_{i1}$ using the CP-ABE algorithm \textsf{Encrypt}:
\begin{displaymath}
\abeEncrypt_{T_{i1}}(\homoEncrypt(\delta|\textbf{x}_i-\textbf{x}_j|^2+\delta')), \abeEncrypt_{T_{i1}}(\homoEncrypt(\delta\tau^2+\delta'))
\end{displaymath}

\STATE
 The doubly nested ciphertexts are sent back to $\aquery_j$, and if $\aquery_j$'s private key $\mathbf{SK}$ satisfies the access tree $T_{i1}$, he can decrypt them and use his Paillier key pair to decrypt the ciphertext again to achieve $\delta|\textbf{x}_i-\textbf{x}_j|^2+\delta'$ and $\delta\tau^2+\delta'$. Then he is able to compare two values to learn whether $dist(\asp_i,\aquery_i)$ is less than, equal to or greater than $\tau$.
\end{algorithmic}
\end{algorithm}

\begin{theorem}\label{theorem:1-in}
If $\aquery$'s distance to $\asp$ is less than $\tau$, 1evel 1 query is equivalent to level 4 query after $O(\log \tau)$ tries.
\end{theorem}
\begin{proof}
For sake of visualization, we prove the theorem in Euclidean plane, but the proof also holds in Euclidean space.
\begin{figure}[!th]
\begin{center}
\scalebox{0.55}{\input{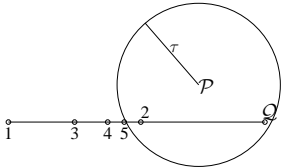}}
\caption{$\hat{\text{x}}$ being inferred by binary search}
\label{fig:binarysearch}
\end{center}
\end{figure}

First draw a circle whose center is $\asp$'s location and the radius is $\tau$. Then, if $\aquery$ is inside this circle, his level 1 query result is `$<$'; if he is outside the circle, the result is `$>$'; if he is just on the circle, the result is `$=$'.

$\aquery$ executes level 1 queries at another random place $\textbf{x}'$ which is $2\tau$ apart from his current location $\textbf{x}$ (i.e., $|\textbf{x}-\textbf{x}'|=2\tau$). Since the radius is $\tau$, $\textbf{x}'$ must be outside the circle. Then, he uses binary search on the line $(\textbf{x}',\textbf{x})$ to find the point $\hat{\textbf{x}}$ such that $|\hat{\textbf{x}}-\textbf{y}|=\tau$ (i.e., the intersection point with the circle). Figure \ref{fig:binarysearch} illustrates this process, where point with number $i$ represents the location where the $i$-th query is executed, and the point $\aquery$ is his initial location. 

The querier repeat the above process by randomly selecting  two more different points ${\textbf x}'$.
We then found three points on the circle. Consequently  the location \textbf{y} is successfully found.
The querier needs at most $\log_2(2\tau)$ tries to find a point on the circle, and three such points are needed to locate \textbf{y}, so \textbf{y} can be calculated after at most $3\log_2(2\tau)$ times for level 1 query.
\end{proof}

\begin{theorem}\label{theorem:1-out}
Suppose $D$ is the greatest possible distance in the location space, if $\aquery$'s distance to $\asp$ is greater than $\tau$, the expected number of level 1 queries after which $\aquery$ achieves $\asp$'s location is $\Omega((D/\tau)^d\log D)$, where $d$ is 2 for Euclidean plane and 3 for Euclidean space.
\end{theorem}

For the simplicity, we only prove for Euclidean plane, but same proof also holds for Euclidean space.
\begin{proof}
$\aquery$ is outside the circle (the one drawn above), so if he finds another location inside the circle, he can determine the location of $\asp$ as proved. Since $\aquery$ does not know where is the circle, he can only randomly choose any location in the location space to execute the level 1 query. The probability of first guess being inside the circle is approximately
\begin{center}
(Size of circle / Size of Euclidean plane) $\approx(\pi\tau^2)/(XY-1)$
\end{center}
where $X$ is the number coordinates in $x$-axis and $Y$ is the number of coordinates in $y$-axis in the Euclidean plane. The approximation comes from the reason that our location system is discrete system with integer coordinates, and $\aquery$'s current location will not be chosen. We can further deduce that at each time, the probability of $i$-th guess being inside the circle is approximately $\frac{(\pi\tau^2)}{XY-i}$. Therefore, the probability that the point inside the circle will be found at $k$-th try is approximately $(1-\frac{\pi\tau^2}{XY-k})^{k-1}\cdot\frac{\pi\tau^2}{XY-k}$. Then, the expected number of tries until the first success is approximately $\sum\limits_{i=1}^{\infty}{k(1-\frac{\pi\tau^2}{XY-k})^{k-1}\cdot\frac{\pi\tau^2}{XY-k}}$.
Then, we have
\begin{displaymath}
\begin{split}
&\sum\limits_{i=1}^{\infty}{k(1-\frac{\pi\tau^2}{XY-k})^{k-1}\cdot\frac{\pi\tau^2}{XY-k}}\\
&>\sum\limits_{i=1}^{\infty}{k(1-\frac{\pi\tau^2}{XY})^{k-1}\cdot\frac{\pi\tau^2}{XY}}\\
&=\frac{XY}{\pi\tau^2}=\Theta((D/\tau)^2)
\end{split}
\end{displaymath}

Therefore, expected number of level 1 queries after which a point inside the circle is guessed is $\Omega((D/\tau)^2)$. After this point is found, the point on the circle can be found using binary search, which leads to $\Theta(log D)$. With three this kind of points, $\asp$'s location can be calculated. Therefore, total expected number of level 1 queries needed to correctly locate $\asp$'s location is $\Omega((D/\tau)^d\log D)$.

Similarly, it can be proved that the expected total number in Euclidean space is $\Omega((D/\tau)^3\log D)$.
\end{proof}

So far, 4 different levels of query protocols are constructed. However, note that level 1-3 queries are equivalent to the level 4 query unless some restrictions are applied, which is proved above. Hence, some restrictions should be applied to protect user's location privacy.

According to Theorem \ref{theorem:3}, during the time period when $\asp$'s location does not change, level 3 query is equivalent to level 4 query unless level 3 queries are limited to three times (two times in Euclidean plane) in this period. Thus, the $\asp$ can choose to discard the query requests after three times of queries.

According to Theorem \ref{theorem:2}, in the level 2 query, information is leaked when one query returns that distance is greater than $\tau$ and another one returns that the distance is less than $\tau$. So, $\asp$ can choose to discard the query requests when the comparison result changes (e.g., from $|\textbf{x}-\textbf{y}|<\tau$ to $|\textbf{x}'-\textbf{y}|>\tau$). Although not responding also leaks some information, this let $\aquery$ learn only that the distance is between two pre-calculated two values.

Similar actions can be taken by $\asp$ in the level 1 query. He responds to queries until the comparison result changes, and not responding to queries let $\aquery$ learn only that the point on the circle is somewhere between two points, and thus protecting $\asp$'s location.

\subsection{Restrictions for $\delta,\delta'$}\label{section:deltas}
As mentioned in Section \ref{section:compare}, $\delta|\textbf{x}-\textbf{y}|^2+\delta'$ and $\delta\tau^2+\delta'$ should be less than the modulo $n$, where $n$ is one of the parameters in Paillier's cryptosystem (Section \ref{section:paillier}). Otherwise, due to the modular operations, the two parameters cannot be compared.

Normally $n=pq$ is a 1024-bit number, which indicates $n>=2^{1023}$. In Euclidean plane, the greatest possible distance in a map of the world is $\sqrt{2}C$, where $C$ is the circumference of the earth (approximately 40000km). This value is approximately equal to $6\cdot 10^7\approx 2^{26}$. Therefore, $|\textbf{x}-\textbf{y}|^2\le 2^{52}$, so it is sufficient to let $\delta\in\mathbb{Z}_{2^{970}}$ and $\delta'\in\mathbb{Z}_{2^{1022}}$. Then, $\delta|\textbf{x}-\textbf{y}|^2+\delta' < 2^{1023} < n$. In Euclidean space, the greatest possible distance is the above distance in a map of the world plus atmosphere height (vector addition). This value is approximately equal to the largest distance above (We estimate the atmosphere height as 32km since 99\% of the air is within it, which is too small  when compared with the circumference of the earth). Therefore, the restrictions to $\delta,\delta'$ remain same.

\section{Performance Evaluation}
\label{sec:evaluation}

In this section, we evaluate the extra communication and computation overhead introduced in our Privacy-preserving Location Query Protocol (PLQP).

Large Number Arithmetic library for smartphone is unavailable currently, so we implemented our protocol in a computer with only one CPU underclocked to 900MHz, whose computation ability is similar to a smartphone. We used GMP library and CP-ABE toolkit \cite{cp-abe} to implement the protocol in Ubuntu 11.04.

Every parameter's length is same as the construction, and we randomly picked two locations for a querier $\aquery$ and a publisher $\asp$. Then, we executed each level query for 1000 times and measured the average running time for each. Since the purpose of the evaluation is to evaluate the computation performance, so we issued a decryption key (of CP-ABE) containing all attributes, which satisfies any access tree, to the querier. In addition, it is well studied in previous works (\cite{jung2013cloud,cp-abe,hasbe}) that encryption and decryption time is proportional to the number of attributes (leaf nodes) in the access tree, so we fixed the attributes in each access tree to ten in every query and did not further analyzed its impact on run time.

\begin{table}[!ht]
\centering \caption{Computation Overhead}
\label{table:computation}
\centering
\begin{tabular}{c|c|c}
\hline \hline
Query Level & $\aquery$'s Run Time (ms) & $\asp$'s Run Time (ms)\\
1 & 577.49 & 919.24 \\
2 & 588.02 & 909.53 \\
3 & 492.89 & 704.85 \\
4 & 413.05 & 702.71 \\
\hline \hline
\end{tabular}
\end{table}
\begin{table}[!ht]
\centering \caption{Communication Overhead}
\label{table:communication}
\centering
\begin{tabular}{c|c|c}
\hline \hline
Query Level & $\aquery\rightarrow\asp$ (Bytes) & $\asp\rightarrow\aquery$ (Bytes)\\
1 & 1280 & 6592 \\
2 & 1536 & 6592 \\
3 & 1280 & 3296 \\
4 & 0 & 3052 \\
\hline \hline
\end{tabular}
\end{table}

Table \ref{table:computation} shows the average run time of each query at the querier's and the publisher's side. We found the run time is dominated by the encryption and decryption algorithms of CP-ABE, and the total run time of each query is less than 1.5 seconds. Also, Table \ref{table:communication} shows that the communication overhead is less than 10 Kilobytes. Note that we only listed the extra overhead in the tables. The total overhead should include other regular overhead (control messages, ACKs etc.). In conclusion, the computation and communication overhead of our protocol is low enough to be used in a real mobile network.

\section{Conclusion}
\label{sec:conclusion}
In this paper, we proposed a fine-grained Privacy-preserving Location Query Protocol (PLQP), which successfully solves the privacy issues in existing LBS applications and provides various location based queries. The PLQP uses our novel distance computation and comparison protocol to implement semi-functional encryption, which supports multi-levelled access control, and used CP-ABE as subsidiary encryption scheme to make access control be more fine-grained. Also, during the whole protocol, unless intended by the location publisher, the location information is kept secret to anyone else. We also conducted experiment evaluation to show that the performance of our protocol is applicable in a real mobile network.



\end{document}